\documentclass{IEEEcsmag}

\usepackage[colorlinks,urlcolor=black,linkcolor=black,citecolor=black]{hyperref}
\expandafter\def\expandafter\UrlBreaks\expandafter{\UrlBreaks\do\/\do\*\do\-\do\~\do\'\do\"\do\-}
\usepackage{upmath,color}
\usepackage{hyperref}
\usepackage{tabularx, booktabs}
\usepackage{enumitem}
\usepackage{xurl}
\usepackage{orcidlink}

\jvol{XX}
\jnum{XX}
\paper{8}
\jmonth{~}
\jname{Computing in Science \& Engineering}
\jtitle{Technology Research Software}
\pubyear{2026}

\providecommand{\doi}[1]{\href{https://doi.org/#1}{#1}}

\newcolumntype{b}{X}
\newcolumntype{m}{>{\hsize=.6\hsize}X}
\newcolumntype{s}{>{\raggedright\hsize=.3\hsize}X}

\begin{document}

\sptitle{Article Type: Research Software Engineering Department}

\title{Technology Research Software: An Often Overlooked Category of Research Software}

\author{Wilhelm Hasselbring \orcidlink{0000-0001-6625-4335}}
\affil{Software Engineering, Kiel University, Kiel, 24098, Germany}

\author{Daniel S. Katz \orcidlink{0000-0001-5934-7525}}
\affil{University of Illinois Urbana-Champaign, Urbana, IL, USA}

\author{Rob van Nieuwpoort \orcidlink{0000-0002-2947-9444}}
\affil{~Leiden University, 2300 RA Leiden, The Netherlands}

\markboth{Research Software Engineering Department}{Research Software Engineering Department}

\begin{abstract}
Research software has been categorized for various goals.
One fundamental dimension of such categorizations is the role that the software plays in the research process. Recently, a new role category has emerged: Technology Research Software, which covers research software developed in technology research. Until now, this category of technology research software has often been overlooked and neglected within the research software engineering community.
In this article, we explain Technology Research Software and its primary sub-roles.
Technology readiness levels are an established method of estimating the maturity of technologies, including software systems. For technology research software, these readiness levels define secondary sub-roles.
To illustrate the concept of technology research software and to make it more tangible, we present examples of research software that, depending on its specific use within or outside of research, take on the role of technology research software as well as that of another research software category.

\end{abstract}

\maketitle

\chapteri Research software is typically developed to meet specific research needs, and often has unique requirements that are different from standard commercial software.
However, research software is gaining appreciation and endorsement for research and as a research result itself.

Research Software Engineering (RSE) is a specialized field that applies software engineering principles to address the unique challenges posed by developing software for scientific and academic research, with the goal of enhancing the efficiency, reproducibility, and impact of research outcomes. 

Recently, a multidimensional categorization of research software, along the dimensions of roles, readiness, developers, and dissemination has been introduced~\cite{Categorization2025}.
This categorization is intended as a basis of institutional guidelines and checklists for research software development. For instance, appropriate software engineering methods for the individual categories may be recommended to meet their specific quality requirements. Categorizations may also be used for a better assessment of existing software when deciding to reuse it. Research funding agencies may use the categories to define appropriate funding schemes. For instance, the German Research Foundation DFG offers a specific funding program for the category of Research Software Infrastructure.\footnote{\url{https://www.dfg.de/en/research-funding/funding-opportunities/programmes/infrastructure/lis/funding-opportunities/research-software-infrastructures}}

When presenting this categorization at various events, it often became apparent that the top-level category of Technology Research Software required some particular explanation, because this is a new and, so far, unfamiliar term. This observation prompted the current department article.

We start with a look at technology research as a research method. Then, we introduce the role-based categorization of
research software, and specifically take a look at the primary and secondary sub roles of technology research software. Technology readiness levels are the basis for these secondary sub roles. Next, we illustrate the relation of the technology research
software category to other categories with some research software examples, before we  conclude with an outlook to future work.

\section{Technology Research as a Research Method}

Simon~\cite{Simon_sciences_1996} distinguishes between two kinds of academic disciplines: the sciences of the natural that study and describe our natural environment, and the sciences of the artificial that prescribe and create artifacts that change our environment. The sciences of the artificial span professional schools, such as engineering and information systems, which primarily design, create, and evaluate artifacts that are useful to society. This is called \textit{Design Science}~\cite{Simon_sciences_1996}. Storey and Baskerville~\cite{Storey2025} discuss the digitalization of the natural sciences (computational science) from a design science perspective.

The ACM Special Interest Group on Software Engineering (SIGSOFT) has defined evidence standards for empirical software engineering research. As a term synonymous with design science, they have defined the \textit{Engineering Research} standard for research involving the invention and evaluation of technological artefacts.\footnote{\url{https://github.com/acmsigsoft/EmpiricalStandards/blob/master/docs/standards/EngineeringResearch.md}}

Van Nieuwpoort and Katz~\cite{NieuwpoortKatz2024} use the term \textit{Technology Research} for this area of study. In technology research, which is most often conducted in computer science, but also in other disciplines, research software usually plays a special role. 
In this context, the research software \textit{itself} is a key research tool.
For instance, it may be a software prototype that demonstrates or explores a novel technological concept. One example is a computer science researcher examining the performance of different programming language design options using a prototype compiler.
In this case, the prototype compiler is research software since it is an artifact produced by and intended for computer science research. We therefore refer to this category of software as \textit{Technology Research Software}.

In the present article, we use the term \textit{Technology Research}, and consider design science and  engineering research to be synonymous. Technology research software is both a means and the \textit{result} of technology research. Conversely, in research fields such as the natural sciences or the humanities, research software is usually only a \textit{means} to obtain research results in the form of answers to original research questions.

\section{Role-Based Categorization of Research Software}
\label{sec:roles}

Research software can be used to collect, process, analyze, and visualize data, as well as to model complex phenomena and run sophisticated simulations. Research software is also developed to control and monitor lab experiments and environmental observations. In technology research, research software constitutes a new paradigm of scientific inquiry next to theory and experiment and acts as a proof-of-concept to invent and evaluate new technological artifacts, including software. Research software also provides the infrastructure to manage, publish, and archive research data and software.
Thus, research software may take various \emph{roles} in the research process.

Van Nieuwpoort and Katz~\cite{NieuwpoortKatz2024} categorize the roles of research software as 
\begin{itemize}
 \item an integral component of instruments used in research,
 \item as the instrument itself,
 \item for analyzing research data,
 \item for presenting research results,
 \item for assembling or integrating existing components,
 \item as infrastructure or an underlying tool, 
 \item for facilitating research-oriented collaboration, and
 \item itself as a research tool for technology research.
\end{itemize}
Based on discussions with the first author of the present article, van Nieuwpoort and Katz extended their categorization, and published Version~2~\cite{NieuwpoortKatz2024} with the missing \textit{Technology Research Software} category, which is the last item in the previous list.

Hasselbring et al.\ \cite{Categorization2025} present a multidimensional research software categorization. Here, we only consider their \textit{role} and \textit{readiness} dimensions. The following three top-level categories are included for the role dimension:
\begin{enumerate}
	\item \emph{Modeling, Simulation, and Data Analytics} of, e.g., physical, chemical, social, linguistic, or biological processes in spatio-temporal contexts.
	\item \emph{Technology Research Software} in technology research.
	\item \emph{Research Infrastructure Software}, such as research data and software management systems.
\end{enumerate}
The assignment of research software to categories may evolve over time. For instance, software specifically developed for a research question (usually Categories 1 \&\ 2) can later turn into infrastructure software (Category~3).
In different contexts, a software may also be in multiple categories at the same time.

\begin{figure*}[htp]
	\includegraphics[width=\textwidth]{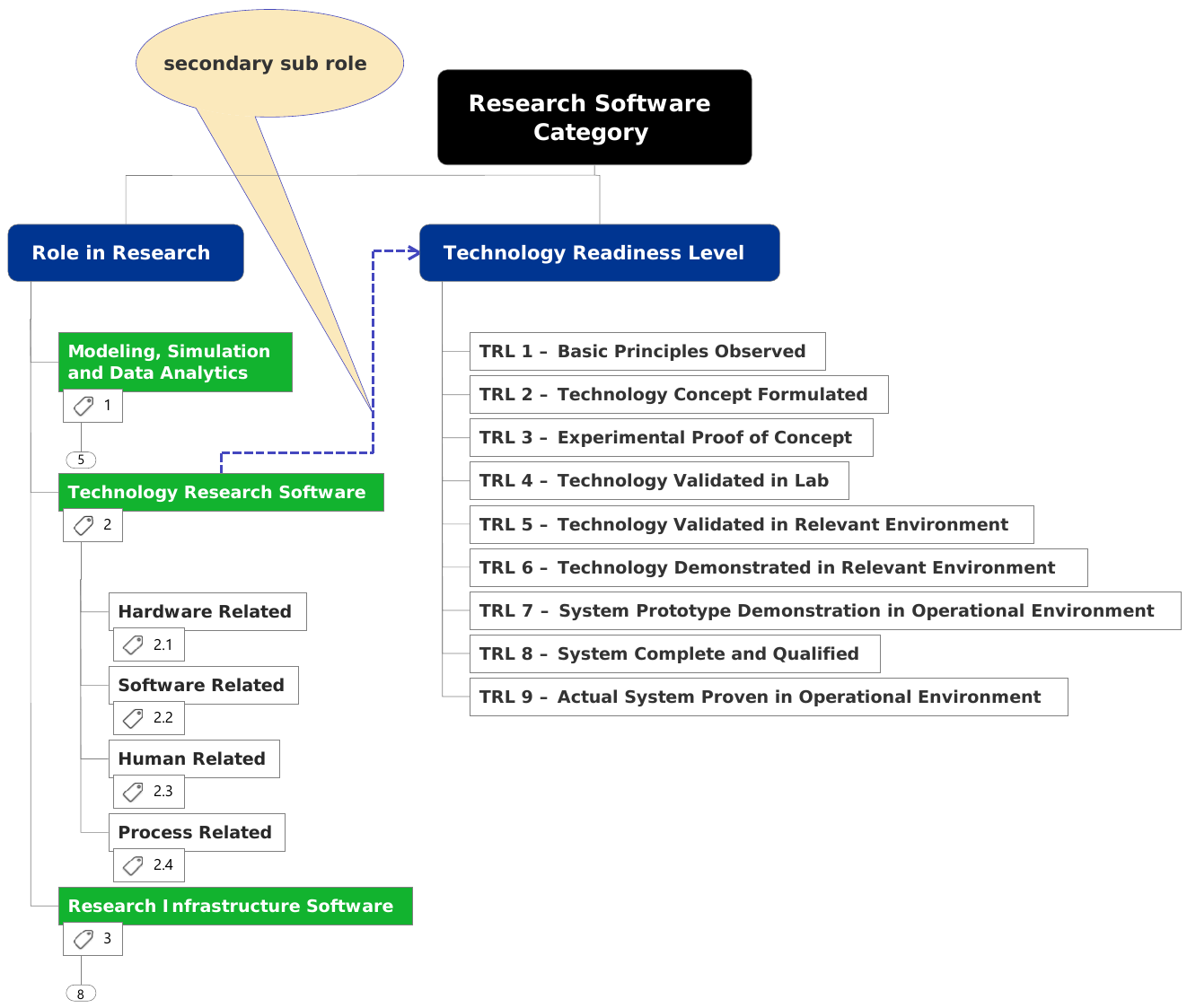}
	\caption{Extract of the multidimensional categorization of research software~\cite{Categorization2025}, restricted to the dimensions of roles for technology research software and the technology readiness.}
	\label{fig:SoftwareCategorization}
\end{figure*}

An extract of the multidimensional categorization~\cite{Categorization2025} with role Category~2 and the readiness dimension is displayed in \autoref{fig:SoftwareCategorization}. For a description of role Categories 1 \&\ 3, refer to~\cite{Categorization2025}.

\section{The Primary Sub Roles of Technology Research Software}
\label{sec:Primary}

Technology research software is developed and used in technology research, and may be related to certain target contexts:

\begin{enumerate}[label=2.\arabic*.]
\itemindent=1em
	\item \textit{Hardware}-related technology research software (usually as embedded software)
    \item \textit{Software}-related technology research software (e.g.,\ as part of an operating system)
    \item \textit{Human}-related technology research software (with a user interface)
    \item \textit{Process}-related technology research software (e.g.,\ as part of business, development or production processes)
\end{enumerate}
These contexts constitute the \textit{primary} sub roles of technology research software, as can be seen on the left of \autoref{fig:SoftwareCategorization}.
Again, one research software may fall into multiple categories. For example, an embedded software with interactive user interface is both hardware-related and human-related.

In the next section, we will examine technology readiness levels for research software. The section after that explains how these technology readiness levels constitute \textit{secondary} sub-roles for technology research software.

\section{Technology Readiness Levels for Research Software}

Technology is the application of conceptual knowledge for achieving practical goals, especially in a reproducible way.  The word \textit{technology} can also mean the products resulting from such efforts, including both tangible tools such as utensils or machines, and intangible ones such as software.

Technology readiness levels (TRLs) are a method for estimating the maturity of technologies. TRLs enable consistent and uniform discussions of technical maturity across different types of technology.

The technology readiness levels may be applied to software~\cite{niemela2004use}:
\begin{description}
	\item[TRL 1] (Basic Principles Observed) A concept that could be
implemented in software.
	\item[TRL 2] (Technology Concept Formulated) Initial software design.
	\item[TRL 3] (Experimental Proof of Concept) Software prototype that validates initial concepts.
	\item[TRL 4] (Technology Validated in Lab) Software integrated with simulated elements as appropriate.
	\item[TRL 5] (Technology Validated in Relevant Environment) The basic software components are integrated with reasonably realistic supporting elements, so that they can be tested in a simulated environment.
	\item[TRL 6] (Technology Demonstrated in Relevant Environment) Software released as ``Beta'' versions.
	\item[TRL 7] (System Prototype Demonstration in Operational Environment) Software demonstrated
as an actual system prototype in an operational environment.
	\item[TRL 8] (System Complete and Qualified) Software has been demonstrated to work in its final form and under expected conditions.
	\item[TRL 9] (Actual System Proven in Operational Environment) Actual application of the software in its final form and under real-world conditions,
\end{description}
These TRLs may be applied to all types of research software, thus, the dimensions are \textit{orthogonal}: every research software may be classified independently in each dimension~\cite{Categorization2025}.

For technology research software, these TRLs can also be interpreted as sub roles, which will be explained in the following section.

\section{The Secondary Sub Roles of Technology Research Software}
\label{sec:Secondary}

\autoref{fig:SoftwareCategorization}, right, shows the resulting readiness-based categorization with the TRL~1 to TRL~9 as categories. 
For technology research software, the TRL titles can be read as \textit{secondary} sub roles. 
One specific technology research software package may take several such sub roles over its lifecycle, with increasing ``readiness''. It may also take several roles at the same time, within different contexts:
\begin{itemize}
    \item In one project, it may serve as ``Experimental Proof of Concept''  (TRL~3);
    \item in another project, it may already serve as a ``Technology Validated in Lab'' (TRL~4).
    \item Eventually, a technology research software package may even become an ``Actual System Proven in Operational Environment'' (TRL~9).
\end{itemize}
Thus, the TRLs constitute \textit{sub roles} of technology research software.

``Readiness'' is top-level in the categorization in \autoref{fig:SoftwareCategorization}, thus it is its own dimension.
We can also categorize the readiness of modeling, simulation, and data analytics software (role category 1) and research infrastructure software (role category 3) with TRLs. For instance, a simulation software may mature to higher TRLs over its lifetime.
If we had put ``Readiness'' directly below ``Technology Research Software'', it would not be its own dimension, thus we added the cross-link from ``Technology Research Software'' to illustrate the additional, secondary sub-role relationship.

\section{Relation of Technology Research Software to Other Categories}

To explain technology research software in more detail, it is also helpful to illustrate the differences to the other role categories.

For instance, the difference between the categories "Modeling and Simulation" and "Technology Research Software" (without consideration of the TRL sub roles) may be illustrated with control engineering research:
\begin{itemize}
	\item As a control engineering researcher, you may build a \textit{simulation} of a control system.
	\item As a control engineering researcher, you may also build an \textit{actual} control system as a new software system. In an automation lab, this researcher may then experiment with this system (not with the simulation of the system). If this system (which is a technology research software) matures, it may reach higher TRLs.
\end{itemize}
In this case, both the simulation and the actual control system are research software in different categories.
The simulation software could even become part of the actual control system, for example to predict system states and inform control actions, effectively turning it into a technology.

A concrete example for technology research software is the Kieker observability and monitoring framework~\cite{Kieker2025}. Kieker provides monitoring, analysis, and visualization support for application-level monitoring and dynamic analysis of software systems, enabling software engineers to understand the behavior of a software system and the reasons behind it. Such an observability framework is integrated as a component into some existing software system to continuously monitor its execution. As such, it is a software-related technology research software (Category 2.2 of \cite{Categorization2025}). In addition, Kieker is used in various research projects for architecture discovery and for performance evaluation~\cite{Kieker2020}. In these research contexts, Kieker is used for software analytics (Category 1.3 of \cite{Categorization2025}). 

Another example is ROAD (the Radio Observatory Anomaly Detector)~\cite{mesarcik_road_2023}, which is an anomaly detection framework for classifying both commonly occurring anomalies as well as detecting unknown rare anomalies that were never seen before. This work started as technology research software, researching AI-based anomaly detection technology, with a TRL of 4. Currently, ASTRON (Netherlands Institute for Radio Astronomy) is working on deploying ROAD inside the LOFAR (Low Frequency Array) radio telescope as operational system health management software; it thus is a part of an operational scientific instrument, moving into Category 3.1 of \cite{Categorization2025}, with a TRL of 7.

A third example for technology research software is the ARCHES (Autonomous Robotic Networks to help Human Societies) framework for research on digital twins, based on the Robot Operating System (ROS)~\cite{JOSS2024}. As such, it is a hardware-related technology research software (Category 2.1 of \cite{Categorization2025}). In addition, ARCHES was used as control software of ocean observation systems on a research vessel in the Baltic Sea~\cite{Barbie2022}. In this research context, ARCHES was used as control and monitoring software (Category 3.1 of~\cite{Categorization2025}).

A concrete research software package may take on several roles over its lifetime and may also fulfill multiple roles simultaneously in different contexts. Table~\ref{table:rules} provides some guidelines for the evolution of research software roles. This list of guidelines is not meant to be exhaustive, but rather to provide examples. As can be seen, such evolutions generally proceed toward Research Infrastructure Software. Table~\ref{table:rules} also gives some examples in the third column. In addition to Kieker and ROAD, which were mentioned previously, additional examples are SUMO (a traffic simulation package, which is used in traffic research and meanwhile also for real-time traffic prediction, see \url{https://eclipse.dev/sumo}), AMUSE (an astrophysics modeling framework, which has been used in many physics research projects and is now also used in other domains, such as ocean modeling, see \url{https://amusecode.org}), and pandas, which started as research software for the quantitative analysis of financial data and is now a popular data analysis package in Python, see \url{https://pandas.pydata.org/about/}.

\begin{table*}[ht]
    \centering
    \begin{tabularx}{\textwidth}{X X r}
    \toprule
        Role Change & Guideline to Advancement & Example\\
    \midrule
       \mbox{Simulation Software} \newline \mbox{~~~~~~~~~$\rightarrow$ Technology Research Software} & The simulation software is incorporated into a technological product. & SUMO\\
       \mbox{Simulation Software} \newline \mbox{~~~~~~~~~$\rightarrow$ Research Infrastructure Software} & Several research projects reuse the simulation software without making essential changes. & AMUSE\\
       \mbox{Analytics Software} \newline \mbox{~~~~~~~~~$\rightarrow$ Technology Research Software} & The analysis software is incorporated into a technological product. & Kieker\\
       \mbox{Analytics Software} \newline \mbox{~~~~~~~~~$\rightarrow$ Research Infrastructure Software} & Several research projects reuse the analytics software without making essential changes. & pandas\\
       \mbox{Technology Research Software} \newline \mbox{~~~~~~~~~$\rightarrow$ Research Infrastructure Software} & Several research projects reuse the technology software without making essential changes. & ROAD \\
    \bottomrule
\\
    \end{tabularx}
    \caption{Guidelines and examples for evolving research software roles.}
    \label{table:rules}
\end{table*}

\section{Conclusion}

Research software categorizations are intended as tools for stakeholders, such as research software engineers and their group, chair, department, or institute leaders, as well as funders, etc. The categorizations may provide these individuals and organizations with a better understanding of the software they are developing or supporting, offering insights into its nature, purpose, and potential impact, or overall portfolio analysis. This knowledge is essential for informed decision-making, adequate resource allocation, and strategic planning within research institutions or funding organizations. 
With this article, we introduce and explain the new and, so far, unfamiliar term \emph{Technology Research Software}.

One essential use case of research software categorizations is its incorporation into guidelines for research software development. For instance, the term Technology Research Software is already used in the German reference guidelines for developing research software~\cite{Muster-Leitlinie2025}.

In the realm of Research Software Engineering (RSE) Research~\cite{RSER2025}, we expect that the research software categorizations provides a framework for classifying research objects, supporting software corpus analyses, and enhancing our understanding of the different types of research software and their properties. This structured approach may aid in organizing and interpreting the vast landscape of research software, contributing to advancements in RSE methodologies and practices. 
With technology research software, we consider the \textit{result} of technology research, which is also known as design science and engineering research.

As future work, we intend to describe more examples of technology research software and their relationship to other research software categories such that we can illustrate the concept of technology research software with a greater breath than it is possible in the short department article.

\def\refname{REFERENCES}

\newpage

\begin{IEEEbiography}{Wilhelm Hasselbring}{\,}is a professor of software engineering at Kiel University, Germany, and an adjunct professor at the University of Southampton, UK. His research interests include software engineering, distributed systems, and open science. Hasselbring received a PhD in computer science from the University of Dortmund. He is a member of the ACM, IEEE Computer Society, and the German Association for Computer Science, at which he is vice chair of the special interest group on research software engineering. Contact him at
\href{mailto:hasselbring@email.uni-kiel.de}{hasselbring@email.uni-kiel.de}.
\end{IEEEbiography}

\begin{IEEEbiography}{Daniel S. Katz}{\,} is Chief Scientist at the National Center for Supercomputing Applications and Research Professor at the University of Illinois Urbana-Champaign. He received the Ph.D. degree in electrical engineering from Northwestern University, Evanston, IL, USA. His interests include software applications, algorithms, and programming in parallel and distributed computing, as well as policy issues such as citation and credit mechanisms and practices associated with software and data, organization and community practices for collaboration, and career paths for computing researchers. He is a senior member of the IEEE and member of the Computer Society Board of Governors, co-founder and current Associate Editor-in-Chief of the Journal of Open Source Software, co-founder of the US Research Software Engineer Association (US-RSE), and co-founder and steering committee chair of the Research Software Alliance (ReSA).
    Contact him at \href{mailto:d.katz@ieee.org}{d.katz@ieee.org}.
\end{IEEEbiography}

\begin{IEEEbiography}{Rob V. van Nieuwpoort} {\,} is professor in efficient computing and eScience at Leiden university, The Netherlands.
His research involves ways in which computer systems can be designed and used more efficiently. He develops new programming models that make the use of large-scale systems simpler, faster, and more energy efficient. His second research interest is eScience. The field of eScience promotes the use of digital technology in research. eScience brings together IT technology, data science, computational science, e-infrastructure and data- and computation-intensive research across all disciplines, from physics to the humanities. He works on bridging fundamental computer science research and its application in exciting scientific disciplines.
Contact him at \href{mailto:r.v.van.nieuwpoort@liacs.leidenuniv.nl}{r.v.van.nieuwpoort@liacs.leidenuniv.nl}.
\end{IEEEbiography}

\end{document}